\documentstyle[twoside,epsf]{article}

\catcode`\@=11
\long\def\@makefntext#1{
\protect\noindent \hbox to 3.2pt {\hskip-.9pt  
$^{{\eightrm\@thefnmark}}$\hfil}#1\hfill}               

\def\thefootnote{\fnsymbol{footnote}}
\def\@makefnmark{\hbox to 0pt{$^{\@thefnmark}$\hss}}    
        
\def\ps@myheadings{\let\@mkboth\@gobbletwo
\def\@oddhead{\hbox{}
\rightmark\hfil\eightrm\thepage}   
\def\@oddfoot{}\def\@evenhead{\eightrm\thepage\hfil
\leftmark\hbox{}}\def\@evenfoot{}
\def\sectionmark##1{}\def\subsectionmark##1{}}

\oddsidemargin=\evensidemargin
\renewcommand{\thefootnote}{\fnsymbol{footnote}}

\newcounter{sectionc}\newcounter{subsectionc}\newcounter{subsubsectionc}
\renewcommand{\section}[1] {\vspace{12pt}\addtocounter{sectionc}{1} 
\setcounter{subsectionc}{0}\setcounter{subsubsectionc}{0}\noindent 
        {\tenbf\thesectionc. #1}\par\vspace{5pt}}
\renewcommand{\subsection}[1] {\vspace{12pt}\addtocounter{subsectionc}{1} 
        \setcounter{subsubsectionc}{0}\noindent 
        {\bf\thesectionc.\thesubsectionc. {\kern1pt \bfit #1}}\par\vspace{5pt}}
\renewcommand{\subsubsection}[1] {\vspace{12pt}\addtocounter{subsubsectionc}{1}
        \noindent{\tenrm\thesectionc.\thesubsectionc.\thesubsubsectionc.
        {\kern1pt \tenit #1}}\par\vspace{5pt}}

\newcounter{appendixc}
\newcounter{subappendixc}[appendixc]
\newcounter{subsubappendixc}[subappendixc]
\renewcommand{\thesubappendixc}{\Alph{appendixc}.\arabic{subappendixc}}
\renewcommand{\thesubsubappendixc}
        {\Alph{appendixc}.\arabic{subappendixc}.\arabic{subsubappendixc}}

\renewcommand{\appendix}[1] {\vspace{12pt}
        \refstepcounter{appendixc}
        \setcounter{figure}{0}
        \setcounter{table}{0}
        \setcounter{lemma}{0}
        \setcounter{theorem}{0}
        \setcounter{corollary}{0}
        \setcounter{definition}{0}
        \setcounter{equation}{0}
        \renewcommand{\thefigure}{\Alph{appendixc}.\arabic{figure}}
        \renewcommand{\thetable}{\Alph{appendixc}.\arabic{table}}
        \renewcommand{\theappendixc}{\Alph{appendixc}}
        \renewcommand{\thelemma}{\Alph{appendixc}.\arabic{lemma}}
        \renewcommand{\thetheorem}{\Alph{appendixc}.\arabic{theorem}}
        \renewcommand{\thedefinition}{\Alph{appendixc}.\arabic{definition}}
        \renewcommand{\thecorollary}{\Alph{appendixc}.\arabic{corollary}}
        \renewcommand{\theequation}{\Alph{appendixc}.\arabic{equation}}
        \noindent{\tenbf Appendix \theappendixc #1}\par\vspace{5pt}}
\newcommand{\subappendix}[1] {\vspace{12pt}
        \refstepcounter{subappendixc}
        \noindent{\bf Appendix \thesubappendixc. {\kern1pt \bfit #1}}
        \par\vspace{5pt}}
\newcommand{\subsubappendix}[1] {\vspace{12pt}
        \refstepcounter{subsubappendixc}
        \noindent{\rm Appendix \thesubsubappendixc. {\kern1pt \tenit #1}}
        \par\vspace{5pt}}

\newcommand{\textlineskip}{\baselineskip=13pt}
\newcommand{\smalllineskip}{\baselineskip=10pt}

\def\abstracts#1#2#3{{
        \centering{\begin{minipage}{4.5in}\baselineskip=10pt\footnotesize
        \parindent=0pt #1\par 
        \parindent=15pt #2\par
        \parindent=15pt #3
        \end{minipage}}\par}} 

\newcommand{\bibit}{\nineit}

\renewenvironment{thebibliography}[1]
        {\frenchspacing
         \ninerm\baselineskip=11pt
         \begin{list}{\arabic{enumi}.}
        {\usecounter{enumi}\setlength{\parsep}{0pt}
         \setlength{\leftmargin 12.7pt}{\rightmargin 0pt} 
         \setlength{\itemsep}{0pt} \settowidth
        {\labelwidth}{#1.}\sloppy}}{\end{list}}

\newcounter{itemlistc}
\newcounter{romanlistc}
\newcounter{alphlistc}
\newcounter{arabiclistc}

\newcommand{\fcaption}[1]{
        \refstepcounter{figure}
        \setbox\@tempboxa = \hbox{\footnotesize Fig.~\thefigure. #1}
        \ifdim \wd\@tempboxa > 5in
           {\begin{center}
        \parbox{5in}{\footnotesize\smalllineskip Fig.~\thefigure. #1}
            \end{center}}
        \else
             {\begin{center}
             {\footnotesize Fig.~\thefigure. #1}
              \end{center}}
        \fi}

\newcommand{\tcaption}[1]{
        \refstepcounter{table}
        \setbox\@tempboxa = \hbox{\footnotesize Table~\thetable. #1}
        \ifdim \wd\@tempboxa > 5in
           {\begin{center}
        \parbox{5in}{\footnotesize\smalllineskip Table~\thetable. #1}
            \end{center}}
        \else
             {\begin{center}
             {\footnotesize Table~\thetable. #1}
              \end{center}}
        \fi}

\def\@citex[#1]#2{\if@filesw\immediate\write\@auxout
        {\string\citation{#2}}\fi
\def\@citea{}\@cite{\@for\@citeb:=#2\do
        {\@citea\def\@citea{,}\@ifundefined
        {b@\@citeb}{{\bf ?}\@warning
        {Citation `\@citeb' on page \thepage \space undefined}}
        {\csname b@\@citeb\endcsname}}}{#1}}

\newif\if@cghi
\def\cite{\@cghitrue\@ifnextchar [{\@tempswatrue
        \@citex}{\@tempswafalse\@citex[]}}
\def\citelow{\@cghifalse\@ifnextchar [{\@tempswatrue
        \@citex}{\@tempswafalse\@citex[]}}
\def\@cite#1#2{{$\null^{#1}$\if@tempswa\typeout
        {IJCGA warning: optional citation argument 
        ignored: `#2'} \fi}}

\def\pmb#1{\setbox0=\hbox{#1}
        \kern-.025em\copy0\kern-\wd0
        \kern.05em\copy0\kern-\wd0
        \kern-.025em\raise.0433em\box0}

\def\fnt#1#2{\footnotetext{\kern-.3em
        {$^{\mbox{\scriptsize #1}}$}{#2}}}

\font\tenrm=cmr10
\font\tenit=cmti10 
\font\tenbf=cmbx10
\font\bfit=cmbxti10 at 10pt
\font\ninerm=cmr9
\font\nineit=cmti9

\font\eightrm=cmr8

\def\qed{\hbox{${\vcenter{\vbox{                        
   \hrule height 0.4pt\hbox{\vrule width 0.4pt height 6pt
   \kern5pt\vrule width 0.4pt}\hrule height 0.4pt}}}$}}

\renewcommand{\thefootnote}{\fnsymbol{footnote}}        

\begin{document}


\normalsize\textlineskip
\thispagestyle{empty}
\setcounter{page}{1}


\rightline{TIFR/TH/00-63}
\rightline{hep-th/0011185}
\vspace*{0.3truein}

\centerline{\bf Stable Non-BPS States and Their Holographic Duals
\footnote{Based on an invited talk given by Sunil Mukhi at Strings
2000, Michigan, July 2000.}}
\vspace*{0.1truein}
\centerline{\footnotesize SUNIL MUKHI and NEMANI V. SURYANARAYANA}
\vspace*{0.015truein}
\centerline{\footnotesize\it Tata Institute of Fundamental Research}
\baselineskip=10pt
\centerline{\footnotesize\it Homi Bhabha Road, Mumbai 400 005, India}

\vspace*{0.2truein}
\abstracts{Stable non-BPS states can be constructed and studied in a
variety of contexts in string theory. Here we review some interesting
constructions that arise from suspended and wrapped branes. We also
exhibit some stable non-BPS states that have holographic duals.
}{}{}

\textlineskip                   
\vspace*{8pt}                  


\def\tilde{\widetilde}
\def\bar{\overline}
\def\to{\rightarrow}
\def\tphi{{\tilde\phi}}
\def\tPhi{{\tilde\Phi}}
\def\cN{{\cal N}}
\def\bN{{\bar N}}
\def\bigone{\hbox{1\kern -.23em {\rm l}}}
\def\ZZ{\hbox{\zfont Z\kern-.4emZ}}
\def\half{{\litfont {1 \over 2}}}
\def\cO{{\cal O}}
\def\tcO{{\tilde {\cal O}}}
\def\hcO{{\hat {\cal O}}}
\def\tcOp{{\tilde {\cal O} p}}
\def\hcOp{{\hat {\cal O} p}}
\def\cOM{{\cal OM}}
\def\cV{{\cal V}}
\def\cF{{\cal F}}
\def\tr{{\rm tr}\,}
\def\hA{{\hat A}}
\def\hcL{{\hat {\cal L}}}
\def\vx{{\vec x}}
\def\vy{{\vec y}}
\def\vX{{\vec X}}
\def\vY{{\vec Y}}
\def\vg{{\vec g}}
\def\vh{{\vec h}}
\def\vH{{\vec H}}
\def\vzero{{\vec 0}}
\def\dbar#1{{\bar {\rm D}#1}}
\def\mofl{(-1)^{F_L}}
\def\mof{(-1)^f}
\def\ddb#1{{${\rm D}#1-{\bar{\rm D}#1}$}}
\def\Dbar{${\bar {\rm D}}$}
\def\onebar{{\bar 1}}
\def\twobar{{\bar 2}}
\def\fbar{{\bar f}}
\def\ftbar{{\bar {f3}}}
\def\crr#1{~{C_{RR}^{(#1)}}}
\def\sign{{\rm sign}\,}
\def\Dp{D$p$}
\def\tq{{\tilde q}}
\def\tick{{\scriptstyle\surd\,}}
\def\rb{\rangle}
\def\lb{\langle}
\def\half{{1\over 2}}
\def\const{{\bg const.}}
\def\det{{\bg det}}
\def\cN{{\cal N}}
\def\darrow{\rightarrow}
\def\tA{{\tilde A}}
\def\tr{{\rm tr~}}
\def\mapdown#1{\Biggl\Downarrow}

\textheight=7.8truein
\setcounter{footnote}{0}
\renewcommand{\thefootnote}{\alph{footnote}}
\hyphenpenalty=8000
\exhyphenpenalty=8000

\section{Introduction}
\noindent
Type II string theory has stable, BPS D$p$-branes with $p=0,2,4,6,8$
in type IIA, and $p=-1,1,3,5,7,9$ in type IIB. For the other values of
$p$ one finds unstable, non-BPS branes: $p=-1,1,3,5,7,9$ in type IIA
and $p=0,2,4,6,8$ in type IIB theory.$^{1,2}$ The spectrum on an unstable
D-brane in superstring theory is the spectrum of a single open string,
but without GSO projection. Hence there is a real tachyon.

The BPS branes are of course stable, while the non-BPS branes can
decay, via tachyon condensation, into the vacuum, or into lower (BPS
or non-BPS) branes. A pair of a BPS brane and its antibrane is also
unstable and can decay similarly.

This is quite a general paradigm. In flat backgrounds, type IIA branes
are either BPS and stable, or non-BPS and unstable. It is interesting
to look for backgrounds which admit non-BPS but stable branes. In this
situation, masses are not protected by BPS formulae. We can hope to
disentangle effects of duality from effects of supersymmetry.

If the backgrounds are themselves non-supersymmetric then things
rapidly become difficult. The most accessible situations are those
where the backgrounds are supersymmetric, but the states that we study
are not. Some examples are orbifolds, orientifolds and Calabi-Yau
compactifications. Another class of examples is provided by suspended
brane constructions.$^{3,4}$ These all have lower supersymmetry than flat
space, which helps to find stable non-BPS states.

In the following, we first investigate brane-antibrane configurations
in the flat-space background of type II superstring theory and
identify some stable non-BPS states. Next, we turn to AdS-type
backgrounds and their holographically dual gauge theories.$^5$ Here,
We analyze stable, non-BPS configurations of branes wrapped over
cycles in the $AdS_5\times T^{1,1}$ background that is dual to
3-branes at a conifold.$^6$ In the course of the discussion we will
make extensive use of the conifold singularity and its
brane-construction dual.$^7$ ALE spaces will also play an auxiliary
role.

\section{Singularities, Brane Duals and Fractional Branes}
\noindent
Let us start with type IIB on a $Z_2$ ALE singularity along the (6789)
directions. Via T-duality along $x^6$, the ALE singularity turns into
a pair of parallel NS5-branes in type IIA string theory, extending
along the (12345) directions and located at different points along
the $x^6$ direction.$^8$

The ALE singularity hides a 2-cycle $\Sigma$ of zero size, which can
be resolved to get an Eguchi-Hanson space. But at the orbifold point,
the NS-NS $B$-field has a flux of ${1\over 2}$ through this
2-cycle.$^9$ In the brane dual, the NS5-branes are symmetrically located
along the $x^6$ circle. This duality extends beyond the orbifold
point. Varying the $B$-flux in the ALE corresponds to varying the
relative $x^6$ separations of the NS5-branes.$^{10}$

If we bring a D3-brane into the plane of an ALE singularity, it can
split into a pair of fractional D3-branes $f3, f3'$ of charge and
tension $\alpha$ and $1-\alpha$ where $\alpha=\int_\Sigma B$ is the
$B$-flux.$^{11}$ The fractional branes are interpreted as: 
\begin{eqnarray}\nonumber
f3:&\quad {\rm D}5~{\rm wrapped~on}~\Sigma\\ \nonumber
f3':&\quad \overline{\rm D}5~{\rm wrapped~on}~\Sigma,\quad\int_\Sigma F =1
\end{eqnarray}

In the dual brane construction, a D4-brane wrapped on $x^6$ can
be brought in to touch the NS5-branes, where it can break into two
pieces (Fig.1(a)).
\begin{figure}[htbp]
\vspace*{13pt}
\centerline{\epsfbox{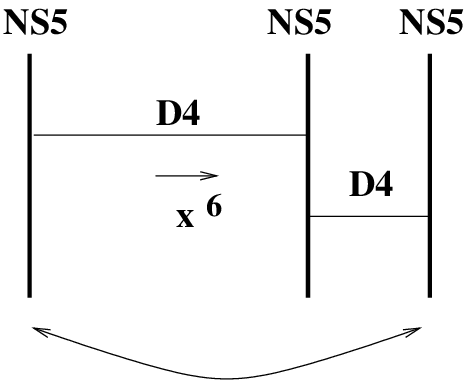}\qquad\qquad\qquad\epsfbox{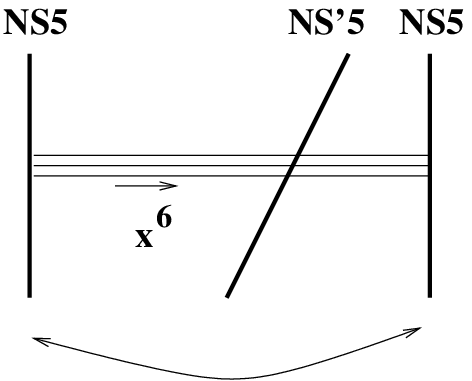}}
\vspace*{13pt}
\fcaption{(a) T-dual of D3-brane at an ALE singularity.\qquad
(b) T-dual of D3-branes at a conifold.}
\end{figure}
The gauge group $U(1)\times U(1)$ and the presence of
bi-fundamental matter is also evident from the brane construction.

An analogous relation holds for the conifold singularity along the
(456789) directions. It is dual to a similar brane construction but
with rotated NS5-branes (Fig.1(b)).$^7$ This model too has bi-fundamental
matter, but also a quartic superpotential as long as there is more
than one D4-brane as shown in the figure.$^{12}$

\section{Fractional Branes and a Stable Non-BPS Configuration}
\noindent

An interesting class of non-BPS brane configurations is obtained
from the system of an adjacent brane-antibrane pair.$^3$ In some
cases, this can be analyzed using perturbative string theory, via
duality to ALE or conifold singularities.

The configuration of interest contains a pair of parallel NS5-branes
oriented as was just discussed. In the two intervals between the 
NS5-branes, we place a D4-brane and a \Dbar4-brane (Fig.2(a)).
\begin{figure}[htbp]
\vspace*{13pt}
\centerline{\epsfbox{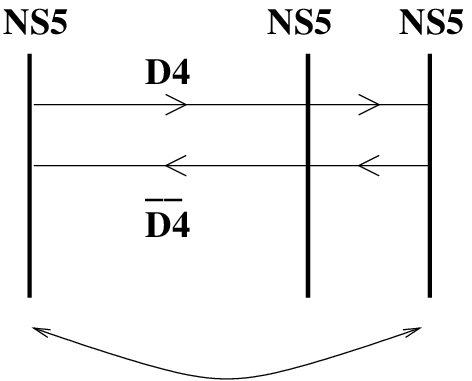}\qquad\qquad\qquad\epsfbox{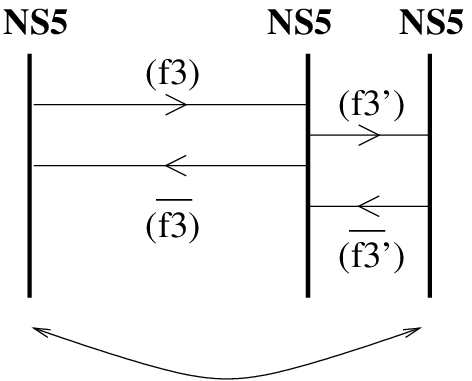}}
\vspace*{13pt}
\fcaption{(a) Brane-antibrane pair at an ALE.\qquad
(b) The four types of (T-dual) fractional branes.}
\end{figure}
The NS5-brane configuration is T-dual to an ALE singularity. The D4
and \Dbar4-brane in the intervals T-dualize into a
fractional D3-brane and a fractional antibrane. Let us try to understand
this correspondence in more detail.

A \ddb3\ pair at a $Z_2$ ALE singularity splits into 
4 distinct types of fractional branes, which we call $f3,f3',\ftbar,\ftbar'$. 
These are interpreted as follows: 
\begin{eqnarray}
f3:\quad &{\rm D}5~{\rm wrapped~on}~\Sigma,\quad \int_{\Sigma}F=0\nonumber\\
f3':\quad &\dbar5~{\rm wrapped~on}~\Sigma,\quad \int_{\Sigma}F=1\nonumber\\
\ftbar:\quad &\dbar5~{\rm wrapped~on}~\Sigma,\quad \int_{\Sigma}F=0\nonumber\\
\ftbar':\quad &{\rm D}5~{\rm wrapped~on}~\Sigma,\quad \int_{\Sigma}F=1
\end{eqnarray}
Introducing a \ddb4\ pair in the brane construction,
we see that it too can break into four distinct pieces:
This is the Coulomb branch, and we can identify the four
fractional branes as in Fig.2(b).

Since we are interested in studying an adjacent \ddb4\ pair, we see
that the dual fractional branes are $f3$ and $\ftbar'$. This system
has a net D5-brane charge of $+2$, and a net D3-brane charge of
$2\alpha -1$.

The open strings connecting adjacent branes correspond in 
the ALE dual to the following Chan-Paton factors: 
\begin{eqnarray}
f3-\ftbar' &:\quad \half(\sigma_3 + i\sigma_2)\otimes
(\sigma_1+i\sigma_2)\nonumber\\
\ftbar' - f3 &:\quad \half(\sigma_3 - i\sigma_2)\otimes
(\sigma_1-i\sigma_2)\nonumber \\
f3'-\ftbar &:\quad \half(\sigma_3 - i\sigma_2)\otimes
(\sigma_1+i\sigma_2)\nonumber \\
\ftbar - f3' &:\quad \half(\sigma_3 + i\sigma_2)\otimes
(\sigma_1-i\sigma_2)
\end{eqnarray}
These are all odd under the ALE projection. Therefore the strings 
connecting $f3$ to $\ftbar'$ have no tachyonic or massless bosonic 
states. In fact, these strings only give massless fermions.

Next we construct the boundary states corresponding to the fractional
D3-branes, and use them to compute the force between the adjacent pair
of interest. There are four independent consistent boundary states for
D3, $\dbar3$, which can be identified with the four fractional branes
$f3, f3', \ftbar', \ftbar$.$^{13}$  
\begin{eqnarray}
\vert {\rm D}3,+ \rangle &= {1 \over 2}\Big( \vert U \rangle_{NSNS}
+ \vert U \rangle_{RR} + \vert T \rangle_{NSNS} +
\vert T \rangle_{RR}\Big) \quad \quad : f3 \nonumber \\
\vert {\rm D}3, - \rangle &= {1 \over 2}\Big(
\vert U \rangle_{NSNS} + \vert U \rangle_{RR}
- \vert T \rangle_{NSNS} - \vert T \rangle_{RR}\Big)
\quad \quad : f3' \nonumber \\
\vert \dbar3, + \rangle &= {1 \over 2}\Big( \vert U \rangle_{NSNS}
- \vert U \rangle_{RR} - \vert T \rangle_{NSNS}
+ \vert T \rangle_{RR}\Big) \quad \quad :
\ftbar' \nonumber \\ \vert \dbar3, - \rangle &= {1 \over 2}\Big(
\vert U \rangle_{NSNS} - \vert U \rangle_{RR} +
\vert T \rangle_{NSNS} - \vert T \rangle_{RR}\Big) \quad \quad :
\ftbar
\end{eqnarray}

The amplitude of interest is:
\begin{eqnarray}
&\int_{0}^{\infty} dl \langle \dbar3, + \vert e^{-lH_c}\vert
{\rm D}3, + \rangle \nonumber = \int_{0}^{\infty} 
{dt \over 2t}\,\tr_{NS-R}\left({1-(-1)^{F} \over 2}\,
{{1-R}\over 2}\,{e^{-2tH_{0}}}\right)
\nonumber \\
&= {v^{(4)}\over 32 (2\pi)^4}\int_0^\infty
{dt\over t^3}\, \big\{
{f_3(\tq)^8 + f_4(\tq)^8 - f_2(\tq)^8\over f_1(\tq)^8}
- 4{f_4(\tq)^4 f_3(\tq)^4 +
f_4(\tq)^4 f_3(\tq)^4\over f_1(\tq)^4 f_2(\tq)^4}\big\}
\end{eqnarray}

This simplifies to:
\begin{equation}
{v^{(4)} \over 16(2\pi)^4}\int_{0}^{\infty}{dt \over t^3}
{f_4(\tq)^8\over f_1(\tq)^8}
\left[1 -4{f_1(\tq)^4 f_3(\tq)^4 \over
f_2(\tq)^4 f_4(\tq)^4}\right]
\end{equation}
The integrand is strictly negative, implying that the force between
the $f3$ and $\ftbar'$ is repulsive. Thus we find that the force
between an adjacent suspended brane-antibrane pair is repulsive.
\begin{figure}[htbp]
\vspace*{13pt}
\centerline{\epsfbox{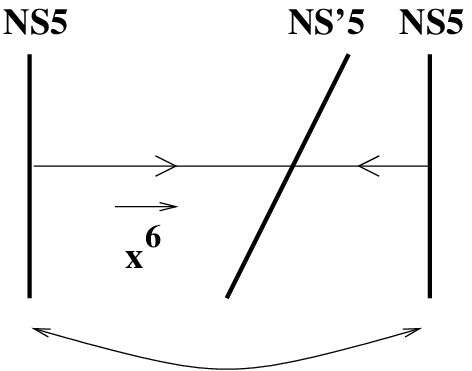}}
\vspace*{13pt}
\fcaption{Adjacent brane-antibrane pair at a T-dual conifold.}
\end{figure}

Now consider a ``twist'' on the configuration of adjacent
brane-antibrane pairs that we discussed earlier.$^4$ We rotate one
NS5-brane as in Fig.3. Thus we now have an NS5 and an NS5'-brane,
making up the brane dual of the conifold. The adjacent D4-brane-antibrane
pair is dual to fractional D3-branes at a conifold.

Physically, we expect a repulsive force between the adjacent brane and
antibrane, as was shown earlier in the unrotated model. But there is
also a classical attraction since the branes cannot separate without
being stretched. This leads to a possibility of stable equilibrium at
finite displacement. In fact we get a more complicated and interesting
result exhibiting a phase transition as a function of the radius $r$
of the compact $x^6$ direction.

The energy of the stretched D4-brane is
\begin{equation}
\cV\, T_4 \sqrt{L^{2}+2r^{2}}
\end{equation}
where $\cV$  is an (infinite) volume factor, $T_4$ is the tension of a BPS 
D4-brane, and $L$ is the separation between the NS5 and NS5'-branes.

We assume that the repulsion is as for the ALE (unrotated) case, since
it comes from strings connecting the \ddb4\ pair across each
NS5-brane. After a calculation,$^4$ we find that the shape of the
potential depends on the separation parameter $L$ as shown in
Figs.4(a) and 4(b) for small $L$ and large $L$ respectively. Here $y
\sim r$ with a constant rescaling.
\begin{figure}[htbp]
\vspace*{13pt}
\centerline{\epsfbox{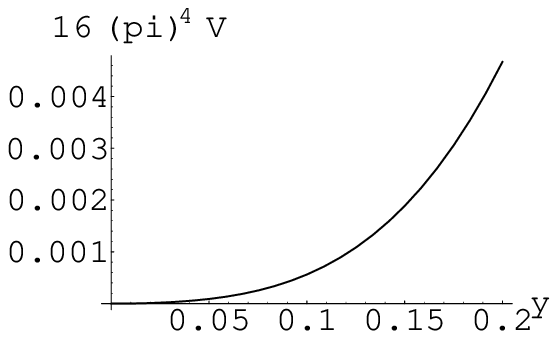}\qquad\qquad\qquad\epsfbox{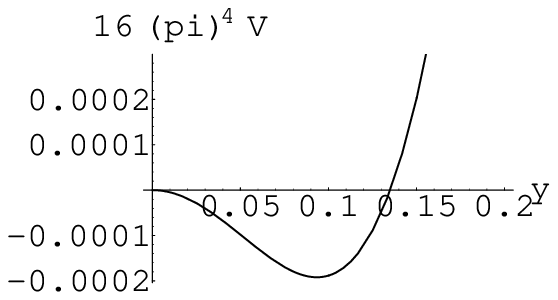}}
\vspace*{13pt}
\fcaption{(a) Brane-antibrane potential for small $L$.\qquad
(b) Brane-antibrane potential for large $L$.}
\end{figure}

Hence the brane and antibrane are aligned for small $L$ but they
separate to a finite distance for large $L$. An estimate shows that 
the potential takes its minimum at $L_c\sim 0.28\, g_s^{-1}$.

\section{Branes at a Conifold and Non-BPS States in $AdS_5$}
\noindent
If we bring $N$ D3-branes to a conifold singularity and take the
large-$N$ limit, we end up with a ${1\over 4}$-supersymmetric
background of type IIB: $AdS_5\times T^{1,1}$ where $T^{1,1}$ is a
particular Einstein 5-manifold.$^6$ If we T-dualize the conifold we get a
model of rotated NS5-branes. $N$ D3-branes at the conifold become $N$
D4-branes wrapped round the $x^6$ circle, as described above.$^7$

The adjacent brane-antibrane model that we have described above does
not have an $AdS$ dual. If we add $N$ D4-branes to it, then the
\Dbar4\ will annihilate against a fractional D4-brane, leaving $N-1$
whole D4-branes plus two fractional D4-branes, as shown in Fig.5.
\begin{figure}[htb]
\vspace*{13pt}
\centerline{\epsfbox{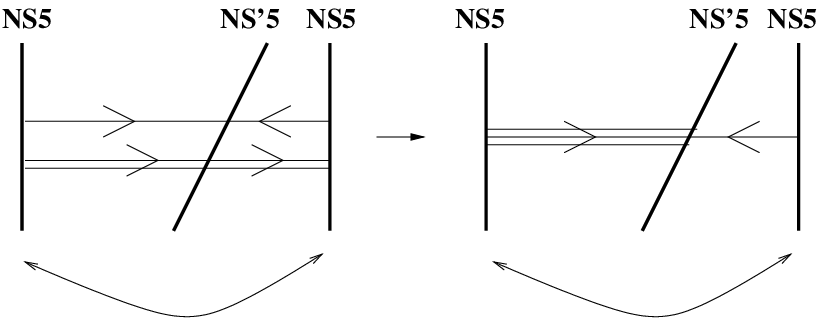}}
\vspace*{13pt}
\fcaption{Adjacent brane-antibrane pair at a T-dual conifold.}
\end{figure}
Let us now describe a stable non-BPS brane construction that, 
instead, does have an $AdS$ dual. 

Take $N$ D4-branes as before and introduce a D2-brane in the first 
interval, as in Fig.6.
\begin{figure}[htb]
\vspace*{13pt}
\centerline{\epsfbox{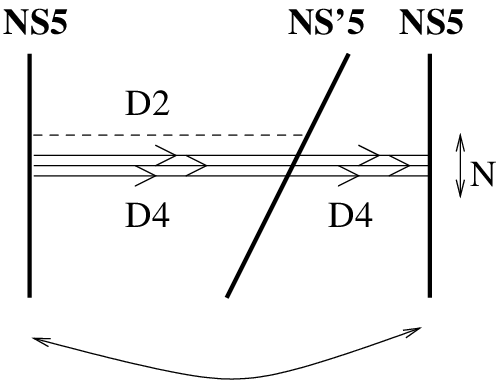}}
\vspace*{13pt}
\fcaption{T-dual conifold with D2 and D4-branes.}
\end{figure}
In the conifold geometry, this corresponds to the introduction of a
fractional D-string in the plane of the singularity. This
configuration is clearly non-supersymmetric. For example, the strings
joining a D2-brane and $N$ D4-branes in the interval will be
tachyonic. This part of the configuration will decay into a stable
bound state of the D4-branes and the D2-brane. While this is BPS by
itself, the neighbouring interval still has only D4-branes, as in
Fig.7. The $(D2,D4)$ bound state and the D4-branes preserve incompatible
supersymmetries.
\begin{figure}[htbp]
\vspace*{13pt}
\centerline{\epsfbox{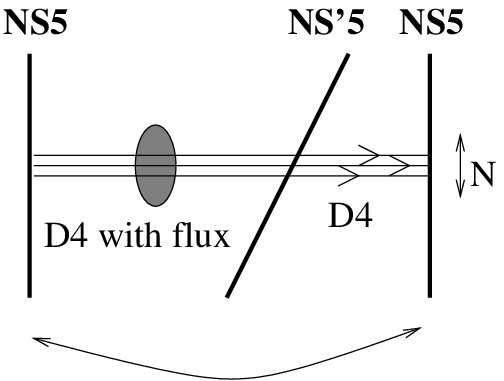}}
\vspace*{13pt}
\fcaption{T-dual conifold with D4-branes and flux.}
\end{figure}
Hence the whole system is non-BPS, much as for an adjacent
brane-antibrane pair. In the conifold geometry, we have a fractional
D-string bound to $N$ $f3$-branes and coincident with $N$ $f3'$
branes.

This system is stable, and we can take the large $N$ limit. In this
limit, the conifold geometry is replaced by its 5-manifold base, the
Einstein space $T^{1,1}$. Topologically, 
$$ 
T^{1,1}\sim S^2 \times S^3
$$ 
The $S^2$ is the same 2-cycle that was of vanishing size before taking
the large-$N$ limit. The fractional D-string was actually a D3-brane
wrapped on this $S^2$.$^{14, 10}$ Hence, in the large $N$ limit, the
fractional D-string can be identified with a ``fat string'' obtained
by wrapping a D3-brane on $S^2$. We will analyze this fat string
further in the following section.

At this point, it is instructive to list all the unwrapped and
wrapped branes of this model:
$$
\matrix{
{\rm Dimension}\hfill\qquad & {\rm Unwrapped}\qquad \hfill & 
S^2 \hfill\qquad & S^3\hfill\qquad & S^2\times S^3\hfill\cr
&&&&\cr 
-1\hfill & D(-1)\hfill & D1\hfill & UD2\hfill & UD4\hfill\cr
\phantom{-}0\hfill & UD0\hfill & UD2\hfill & D3\hfill & D5\hfill\cr
\phantom{-}1\hfill & D1\hfill & D3\hfill & UD4\hfill & UD6\hfill\cr
\phantom{-}2\hfill & UD2\hfill & UD4\hfill & D5\hfill & D7\hfill\cr
\phantom{-}3\hfill & D3\hfill & D5\hfill & UD6\hfill & UD8\hfill\cr
\phantom{-}4\hfill & UD4\hfill & UD6\hfill & D7\hfill & D9\hfill\cr}
$$ 
In this table, the prefix ``U'' indicates an unstable brane. The
remaining branes are stable. Some of these wrapped branes have been
studied previously.$^{14, 10}$ For example, the D5 wrapped on $S^2$ is
known to be a domain wall that augments the gauge group: $U(N)\times
U(N) \rightarrow U(N+1)\times U(N)$, while the D3 wrapped on $S^2$
is our fat string. We would like to understand its holographic dual
description. The Euclidean D-string wrapped on $S^2$ gives rise to an
instanton, while the (unstable) $UD2$ on $S^2$ is an unstable
D0-brane\footnote{These are distinct from the standard type IIB
D-instantons and D0-branes, whose holographic duals have been studied
previously.$^{15,16}$}. We will comment on their holographic duals too.

\section{Properties of the Fat String}
\noindent
The nature of the fat string depends on the B-flux through
$S^2$. In general we have 
$$
\int_{S^2} B_{NS,NS} = \alpha,\qquad \int_{S^2} B_{RR} = \beta
$$
The $U(N)\times U(N)$ gauge theory on the 3-branes has couplings 
and $\theta$-angles given by:$^{17}$
\begin{eqnarray}
\tau_1 &=& \beta + \alpha\tau_s\nonumber \\
\tau_2 &=& -\beta + (1-\alpha)\tau_s
\end{eqnarray}
where $\tau_s = {\chi_{RR}\over 2\pi} + {i\over g_s}$.

The fat string carries D-string charge $\alpha$ and F-string charge
$\beta$, by virtue of the Chern-Simons coupling 
$$
\int B_{NS,NS}\wedge B_{RR}\rightarrow \alpha\int B_{RR} + \beta\int
B_{NS,NS}
$$
on a D3-brane. It is convenient to choose $\beta=0$. The tension of
the fat string can be estimated from integrating the DBI action of a
D3-brane over $S^2$:
$$
T_{\rm fat}\sim T_3\int_{S^2} \sqrt{{\rm det} g + (B_{NS,NS})^2 }
$$
In the flat space limit, the $S^2$ is of zero size and this becomes
$$
T_{\rm fat}\sim T_3\,\alpha
$$
which shows that it is BPS. On the other hand at large
$N$ the dominant contribution comes from
$$
T_{\rm fat}\sim T_3\int_{S^2}\sqrt{g} \sim {N\over (g_s N)^\half \alpha'}
$$

As with fractional branes, there are really two complementary fat
strings, the second one being an anti D3-brane wrapped over $S^2$ and
having a magnetic flux $\int F=1$ over the cycle. We call this a
fat$'$ string. It has a D-string charge $(1-\alpha)$. The non-BPS
nature of fat strings, and their charges, imply that a fat string and
a fat$'$ string can annihilate with loss of energy into a D-string.

Recall how a D-string is understood in holography.$^{18}$ In
$AdS_5\times S^5$, a D-string parallel to the boundary corresponds to
a magnetic flux tube. As the string falls towards the horizon, the
flux tube fattens and in the limit becomes a constant flux. The same
result holds for a D-string in $AdS_5\times T^{1,1}$, but the flux is
in the diagonal of the $U(N)\times U(N)$ gauge group.

The fat string is similarly a flux tube in the boundary theory, but
this time the flux is only in one $U(N)$ factor. This is
consistent with its non-BPS nature. On a 3-brane we have nonlinearly
realized supersymmetry that acts on the gauginos as: 
$$
\delta^*\lambda^{(1)}_\alpha  = {1\over
4\pi\alpha'}\eta^*_\alpha,\quad
\delta^*\lambda^{(2)}_\alpha = {1\over 4\pi\alpha'}\eta^*_\alpha
$$
and linearly realized supersymmetry:
$$
\delta\lambda^{(1)}_\alpha = F^{(1)}_{23}
\sigma^{23\,\beta}_\alpha\,
\eta_\beta,\quad
\delta\lambda^{(2)}_\alpha  = 
F^{(2)}_{23} \sigma^{23\,\beta}_\alpha\,\eta_\beta
$$
If and only if the fluxes are diagonal: $F^{(1)} = F^{(2)}= F$, there
is a surviving set of linearly realized supersymmetries, described by
choosing
$$
\eta^*_\alpha = - 4\pi\alpha' F_{23}
\sigma^{23\,\beta}_\alpha\,\eta_\beta
$$
For non-diagonal fluxes, no supersymmetry is preserved.

One can also compute the potential experienced by the non-BPS fat
string, and also study Wilson/'t Hooft loops in the $AdS$
context.$^{19}$

Let us comment briefly on some of the other wrapped branes. The
euclidean D1 wrapped on $S^2$ is a new ``D-instanton''. It is expected
to be dual to a Yang-Mills instanton in the first factor of
$U(N)\times U(N)$.  It has its own associated sphaleron, the
D2-brane of type IIB wrapped on $S^2$. The relation between the two is
parallel to the one between unwrapped D-instantons and D0-branes,
studied recently.$^{20,16}$

\section{Conclusions}
\noindent
The stable brane-antibrane construction that we have exhibited should
describe an interesting non-SUSY model field theory. Microscopically it
has a pair of branes separated by a finite calculable distance. Such
constructions might be useful in making brane-world type models. It
would be interesting to understand the spectrum and interactions of
the effective low-energy field theory on these branes, which have so
far not been worked out in complete detail.

Recall that the conventional BPS brane constructions are most useful
when we can use S-duality (in type IIB) or the duality with M-theory
(in type IIA). What do we learn from these dualities about
brane-antibrane constructions?  A stable non-BPS brane configuration
in type IIA theory, such as the one we have exhibited, must have a
well-defined M-theory limit as an M5-brane wrapping a 2-cycle. Because
the configuration is not BPS, the 2-cycle will not be holomorphic, so
a novel approach would be required to determine it.

In the present discussion, ``fat'' objects were associated to one
$U(N)$ factor while ``thin'' objects are diagonal in $U(N)\times
U(N)$. This is quite general. The study of branes at more complicated
singularities gives rise to product gauge groups involving
many factors, and one should then be able to construct large numbers
of non-BPS objects associated to one or more of these factors. Each
one will be a ``fat'' object, related to a fractional brane at the
generalized singularity.

The stable non-BPS configurations described here are particularly
suitable for investigation using the AdS/CFT correspondence.
This should help us to generalize many of the notions of holography to
situations without supersymmetry, nevertheless retaining some control
over the dynamics.

\end{document}